\documentclass[twocolumn,superscriptaddress,amsmath,amssymb,aps,prl,10pt]{revtex4-1}

\usepackage{graphicx}
\usepackage{hyperref}

\begin{document}

\title{Universal non-resonant absorption in carbon nanotubes.}

\author{Fabien Vialla}
\affiliation{Laboratoire Pierre Aigrain, \'Ecole Normale Sup\'erieure, CNRS, Universit\'e Pierre et Marie Curie, Universit\'e Paris Diderot, 24 rue Lhomond 75005 Paris, France}
\author{Ermin Malic}
\affiliation{Technische Universit\"at Berlin, Department for Theoretical Physics, Germany}
\author{Benjamin Langlois}
\affiliation{Laboratoire Pierre Aigrain, \'Ecole Normale Sup\'erieure, CNRS, Universit\'e Pierre et Marie Curie, Universit\'e Paris Diderot, 24 rue Lhomond 75005 Paris, France}
\author{Yannick Chassagneux}
\affiliation{Laboratoire Pierre Aigrain, \'Ecole Normale Sup\'erieure, CNRS, Universit\'e Pierre et Marie Curie, Universit\'e Paris Diderot, 24 rue Lhomond 75005 Paris, France}
\author{Carole Diederichs}
\affiliation{Laboratoire Pierre Aigrain, \'Ecole Normale Sup\'erieure, CNRS, Universit\'e Pierre et Marie Curie, Universit\'e Paris Diderot, 24 rue Lhomond 75005 Paris, France}
\author{Emmanuelle Deleporte}
\affiliation{Laboratoire Aim\'e Cotton, \'Ecole Normale Sup\'erieure de Cachan, Universit\'e Paris Sud, CNRS, 91405 Orsay, France}
\author{Philippe Roussignol}
\affiliation{Laboratoire Pierre Aigrain, \'Ecole Normale Sup\'erieure, CNRS, Universit\'e Pierre et Marie Curie, Universit\'e Paris Diderot, 24 rue Lhomond 75005 Paris, France}

\author{Jean-S\'ebastien Lauret}
\affiliation{Laboratoire Aim\'e Cotton, \'Ecole Normale Sup\'erieure de Cachan, Universit\'e Paris Sud, CNRS, 91405 Orsay, France}

\author{Christophe Voisin}
\email{christophe.voisin@lpa.ens.fr}
\affiliation{Laboratoire Pierre Aigrain, \'Ecole Normale Sup\'erieure, CNRS, Universit\'e Pierre et Marie Curie, Universit\'e Paris Diderot, 24 rue Lhomond 75005 Paris, France}

\date{\today}

\begin{abstract}
Photoluminescence excitation measurements in semi-conducting carbon nanotubes show a systematic non-resonant contribution between the well known excitonic resonances. Using a global analysis method, we were able to delineate the contribution of each chiral species including its tiny non-resonant component. By comparison with the recently reported excitonic absorption cross-section on the $S_{22}$ resonance, we found a universal non-resonant absorbance which turns out to be of the order of one half of that of an equivalent graphene sheet. This value as well as the absorption line-shape in the non-resonant window is in excellent agreement with microscopic calculations based on the density matrix formalism. This non-resonant absorption of semi-conducting nanotubes is essentially frequency independent over 0.5~eV wide windows and reaches approximately the same value betweeen the $S_{11}$ and $S_{22}$ resonances or between the $S_{22}$ and $S_{33}$ resonances. In addition, the non-resonant absorption cross-section turns out to be the same for all the chiral species we measured in this study. From a practical point of view, this study puts firm basis on the sample content analysis based on photoluminescence studies by targeting specific excitation wavelengths that lead to almost uniform excitation of all the chiral species of a sample within a given diameter range. 
\end{abstract}

\maketitle


In contrast to graphene, single-wall carbon nanotubes (SWNTs) show marked resonances in their optical spectrum that primarily reflect the one-dimensional quantum confinement of carriers. These resonances that combine one-dimensional and excitonic characteristics have been extensively investigated and are widely used as finger prints of the $(n,m)$ species \cite{Bachilo2002}. However, spectroscopic studies reveal that the absorption of nanotubes does not vanish between resonances and consists of a wealth of tiny structures, such as phonon side-bands, crossed excitons ($S_{ij}$), or higher excitonic states \cite{Plentz2005, Berciaud2010, Miyauchi2006a, Lefebvre2008}. In ensemble measurements, the non-resonant absorption is even more congested due to the contribution of residual catalyst or amorphous carbon and due to light scattering \cite{Naumov2011}. In total, a relatively smooth background showing an overall increase with photon energy is observed, from which it is challenging to 
extract any quantitative information.

 In this study, we show that thorough photoluminescence excitation (PLE) measurements yield a much finer insight into the non-resonant absorption of carbon nanotubes, that reveals the universal features of light-matter interaction in carbon nano-structures \cite{Nair2008}. In particular, we show that the non-resonant absorption of SWNTs per unit area well above the $S_{11}$ or $S_{22}$ resonances reaches an universal value of 0.013$\pm0.003$ in good agreement with the value $\alpha \sqrt{3}$ (where $\alpha$ is the fine structure constant) predicted by a simple band-to-band theory.

\begin{figure}[t!]
	\centering
		\includegraphics[width=0.47\textwidth]{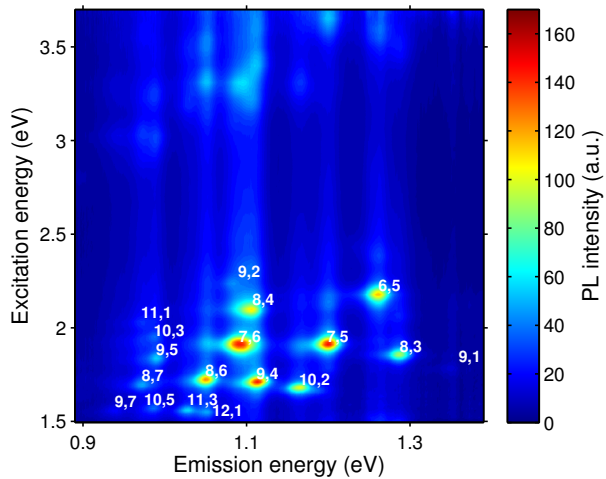}
	\caption{Photoluminescence map of a micelle suspension of HiPCO nanotubes in D$_2$O. The $(n,m)$ indices are assigned to each PL resonance according to the scheme of Ref. \onlinecite{Bachilo2002}. The spots at the bottom of the map correspond to the excitation on the $S_{22}$ resonances of the nanotubes whereas the spots in the upper part of the map correspond to the excitation on the $S_{33}$ levels. \label{fig:PLEmap}}
\end{figure}

Our study of non-resonant absorption is based on the global analysis of PLE maps of ensembles of carbon nanotubes that allows us to deconvolute the contribution of each $(n,m)$ species while keeping a high signal to noise ratio. The sample consists of micelle suspensions (sodium cholate) of HiPCO and CoMoCat nanotubes in deutered water (D$_2$O). All measurements are done at room temperature. The sample is excited by using a cw Xe lamp filtred by a monochromator that provides 5~nm-wide excitation steps throughout the near UV, visible and near IR regions. The luminescence of the suspension is collected in a 90 degree geometry and dispersed in a 30~cm spectrograph coupled to an InGaAs linear detector. The spectra are corrected from the grating and detector efficiencies and are normalized to the incoming photon flux. The PLE map of a HiPCo sample is displayed in Figure~\ref{fig:PLEmap}. It consists of bright spots that correspond to the resonant $S_{22}$ excitation of the $S_{11}$ luminescence of the 
semi-conducting species. The $(n,m)$ species were identified according to the scheme developed by Bachilo \textit{et al.} \cite{Bachilo2002}. These spots are surrounded by a weak but non vanishing background showing up as vertical light-blue strips in the map. In contrast to absorption measurements, the presence of such a background in a luminescence measurement shows that there is an actual absorption by the very species that give rise to the IR luminescence, that is the semi-conducting nanotubes themselves. Due to the luminescence detection scheme, this background can be unambiguously distinguished from contributions such as elastic light scattering, absorption by metallic species, by catalyst or amorphous carbon particles, etc. that blur the non-resonant response of the sample in linear absorption measurements.

\begin{figure}[b!]
	\centering
		\includegraphics[width=0.47\textwidth]{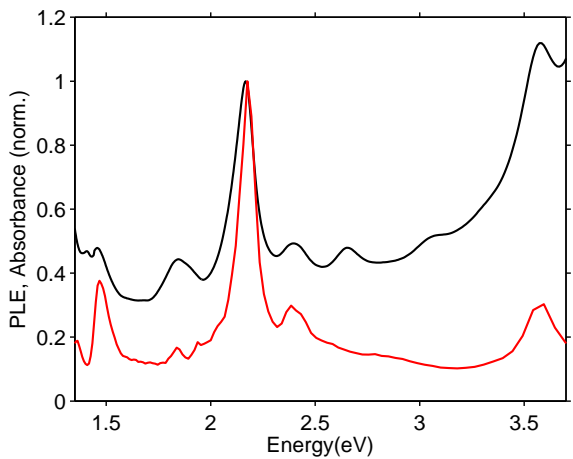}
	\caption{Linear absorption spectrum (black) and PLE spectrum (red) of a (6,5) enriched suspension. Both spectra are normalized to the maximum of the $S_{22}$ resonance to show their respective background contributions. \label{fig:compAbs-PLE}}
\end{figure}

This point is illustrated in Figure~\ref{fig:compAbs-PLE} where the linear absorption spectrum of a chirality enriched (6,5) sample (NanoIntegris) is compared to the PLE spectrum of the same sample. Both spectra are normalized to the maximum of the $S_{22}$ resonance in order to compare their background levels. The latter appears to be 2 to 3 times lower in the case of the PLE spectrum, showing that extrinsic contributions in the linear absorption spectrum are sizable even for sorted samples. 

\begin{figure}[h!]
	\centering
	\includegraphics[width=0.50\textwidth]{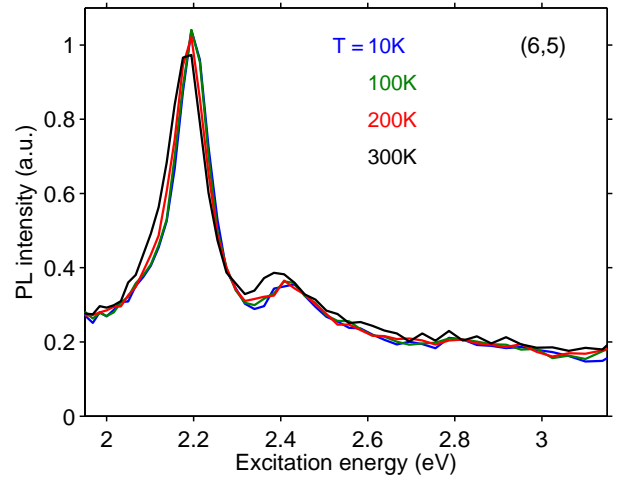}
	\caption{Normalized PLE spectra of the (6,5) species measured at several temperatures. \label{fig:temperature}}	
\end{figure}

Importantly, PLE and absorption spectra can be directly compared if the relaxation from the upper levels down to the light-emitting level is much faster than the competing relaxation processes \cite{Bastard1988}. Several time-resolved studies have shown that non-resonant excitation of SWNTs leads to extremely fast (sub-ps) relaxation to the lower bright level \cite{Lauret2003a, Manzoni2005, koehler13}. We also checked with two-color pump-probe measurements that the population build-up time on the $S_{11}$ level does not depend on the pump wavelength (Appendix, Figure~\ref{fig:femto}). To further make sure that this internal conversion is faster than the competing relaxation processes, we varied the latter by changing the temperature of the sample (from 10 to 300K) \cite{Berger2007} and checked that the PLE spectrum remained unchanged (Appendix, Figure~\ref{fig:temperature} ). In total, we can safely assume that non-resonant PLE measurements accurately reflect the absorption properties of SWNTs.

Due to the numerous chiral species present in the sample and due to the partial spectral overlap of their IR emission lines, a specific analysis method is required in order to quantitatively assess the non-resonant absorption of each species. To this end, we developed a global fitting procedure: When the luminescence is excited at the $S_{22}$ resonance of a given $(n,m)$ species, the corresponding $S_{11}$ emission line becomes the dominant feature of the luminescence spectrum, thus making possible to determine accurately the emission energy and the width. We performed such preliminary fitting for each $(n,m)$ species emission line. From this set of $S_{11}$ energies and widths, we constructed a model function made of the sum of Lorentzian profiles with free amplitudes and fixed energies and widths. Finally, the PL spectrum obtained for each excitation wavelength was fitted to the model function (Appendix, Figure~\ref{fig:globalanalysis}). The resulting amplitude of each Lorentzian component defines the PLE 
amplitude of the corresponding $(n,m)$ species at this excitation wavelength \footnote{
Note however, that this method is not applicable for the species for which the emission energy difference is much smaller than their emission linewidth, giving completely overlapping contributions.}.



\begin{figure}[t!]
	\centering
	\includegraphics[width=0.50\textwidth]{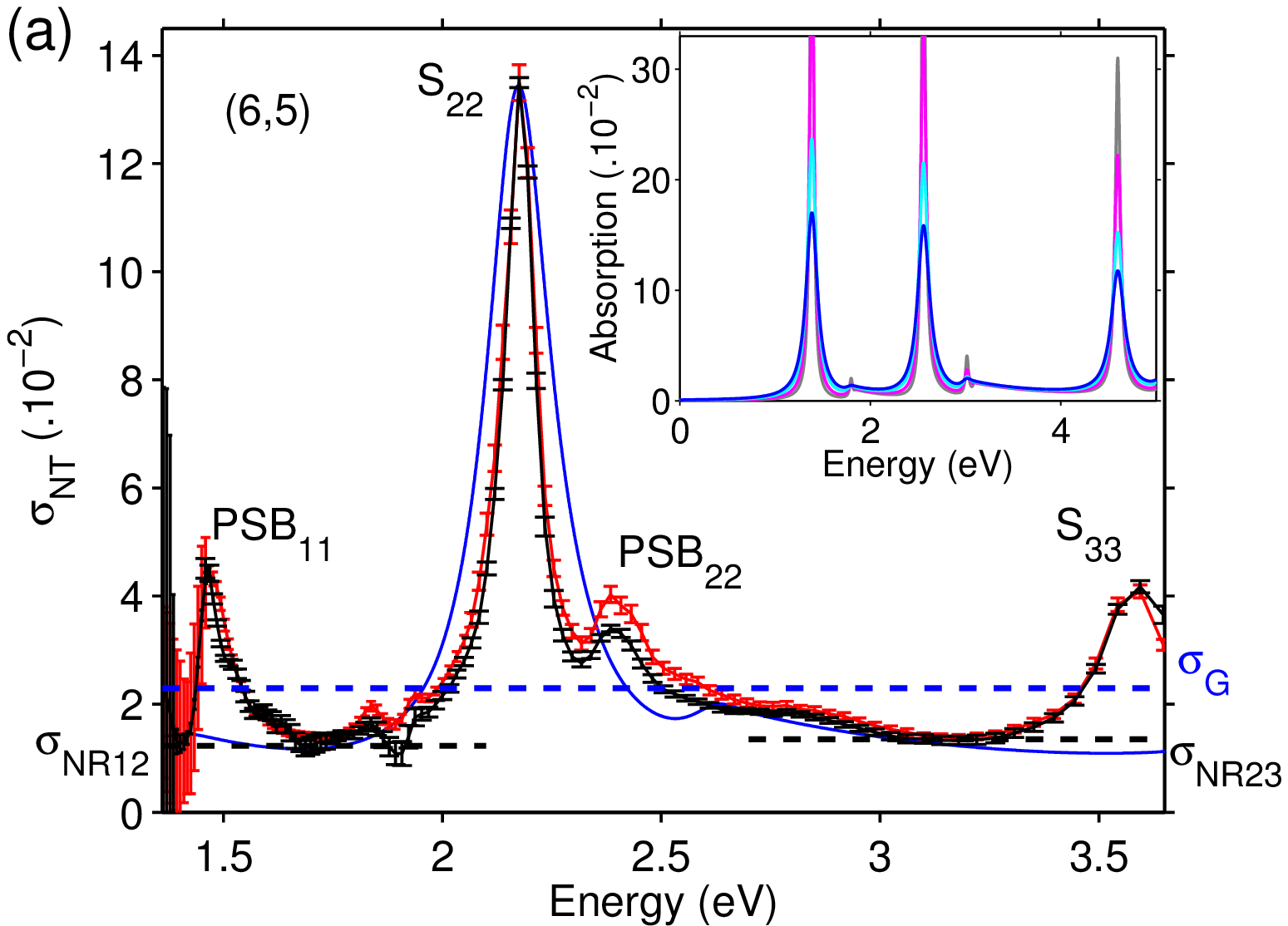}
	\includegraphics[width=0.5\textwidth]{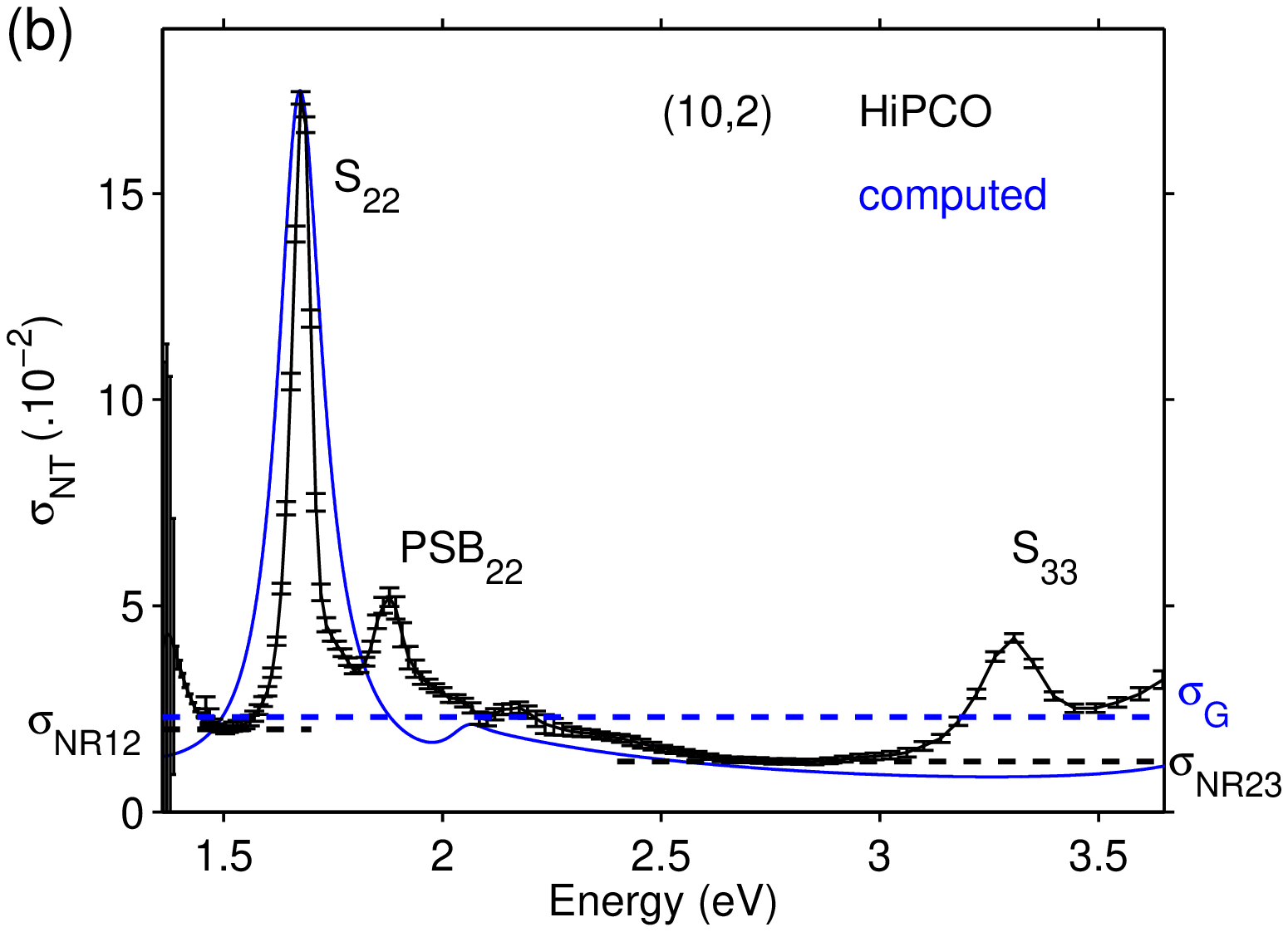}
	\caption{Deconvoluted PLE spectra for the (6,5) (a) and (10,2) (b) species (HiPCO (black) and CoMoCat (red)) extracted from the PL map of Figure~\ref{fig:PLEmap}. The absolute value of $\sigma$ is retrieved from the knowledge of the resonant $S_{22}$ absorption cross-section \cite{Vialla2013} and assuming a flat internal conversion rate. The values are expressed per unit area allowing a direct comparison with the graphene case. The blue dashed line represents the equivalent absorption of a graphene layer. The black dashed lines show the non-resonant absorption plateaus that make up the input of Figure~\ref{fig:HRSvsD}. The solid blue line is the microscopic calculation of the absorption spectrum of the corresponding species after applying a global energy shift and a global convolution with a 150~meV (a) or 130~meV (b) broad Lorentzian. The inset in (a) shows the calculated spectra for a set of broadenings of 70, 100, 150 and 200 meV, showing that the magnitude and the shape of the non-resonant absorption is 
hardly altered. \label{fig:PLEspectrum}}.
\end{figure}

Such a deconvoluted PLE spectrum is displayed in Figure~\ref{fig:PLEspectrum} for two exemplary species. The most prominent features are the $S_{ii}$ resonances (essentially the $S_{22}$ and $S_{33}$ ones) and the associated phonon side-bands \cite{Berciaud2010, Plentz2005}. An additional weaker resonance is observable roughly 0.5~eV above the $S_{22}$ resonance that might be attributed to the $S_{23}$ cross polarized transition \cite{Miyauchi2006a}. In addition to these features, a background is obtained between the excitonic resonances in the PLE measurements. This background is readily observable as a plateau in the 0.5~eV wide energy window free of any resonant feature on the red side of each excitonic resonance. It represents roughly 15\% of the $S_{22}$ absorption cross-section. Interestingly, we can compare the deconvoluted PLE spectrum of a given species (namely the (6,5) one) for two type of samples (black (for HiPCO) and red (for CoMoCat) lines in Figure~\ref{fig:PLEspectrum}~(a)). Although these 
samples differ 
in both the growth method and the post-growth processing, the PLE spectra are remarkably similar, confirming that the non-resonant signal has an intrinsic origin. 

In order to be more quantitative, we make use of our recent measurement of the absolute value of the $S_{22}$ absorption cross-section of most of the chiral species present in this sample \cite{Vialla2013} to rescale the PLE axis into absorption cross-section. In order to take into account the total number of carbon atoms involved in the absorption for each species, we further calculate the (unitless) absorption cross-section per unit area by dividing the absorption cross-section by $\pi d_t$ where $d_t$ is the tube diameter. In addition, this approach provides a direct comparison with the case of graphene that shows a frequency independent absorption cross-section of $\pi \alpha = 0.023$ in the visible and near IR region \cite{Nair2008}.

The non-resonant contribution of SWNTs is defined as the minimum of the absorption cross-section between two resonances, which practically corresponds to a 500~meV wide plateau on the red side of each $S_{ii}$ resonance. We note that depending on the chiral species, trigonal warping effects can spread the transition energies leading  to wider observation windows for the non-resonant signal (between the $S_{22}$ and $S_{33}$ transitions for type II species for instance). Conversely, for large diameter nanotubes, the limited energy splitting between resonances (that may become comparable to their width) may compromize the observation of the plateau or warp its value.

The observation of this non-resonant background is consistent with former PLE measurements \cite{Lefebvre2008, Kimoto2013} and with
recent absorption studies conducted for individual large diameter nanotubes between higher transitions \cite{Blancon2013, Liu2014} showing systematically a sizable background between higher transitions. These studies are in agreement with our observations both regarding the profile and the magnitude of the non-resonant signal. 

We were able to quantitatively reproduce this non-resonant absorption spectrum within microscopic calculations based on the density matrix formalism combined with tight-binding wave functions \cite{malic13}.
The calculation of the frequency-dependent absorption coefficient $\alpha(\omega)$ requires the knowledge of the microscopic polarization  $p_k(t)=\langle a^+_{vk} a^{\phantom{+}}_{ck}\rangle$, which is a measure for the
transition probability between the conduction and the valence band at the wave vector $k$ \cite{malic13}. Applying the Heisenberg equation of motion and exploiting the fundamental commutator relations, we obtain the corresponding semiconductor Bloch equation for $p_k(t)$ within the limit of linear optics \cite{malic13} :
\begin{equation}
\label{bloch}
\dot{p}_{{k}}(t)=-i\Delta\tilde{\omega}_{{k}} p_{{k}}(t) +i \tilde{\Omega}_{
k}(t)-\gamma  p_{{k}}(t).
\end{equation}
The dephasing of the microscopic polarization is taken into account by the  
parameter $\gamma$, which determines the line width of
optical transitions. 
 The Coulomb interaction is  considered within the screened Hartree-Fock level and leads to 
 a renormalization of the transition energy $ \Delta\varepsilon_{k}=\hbar(\omega_c({k})-\omega_v({k}))$ due to the repulsive
electron-electron interaction $W_{e-e}({k},{k'})$ 
\begin{equation}
\label{ren}
\hbar\Delta\tilde{\omega}_{{k}}=\Delta\varepsilon_{k}-\sum_{{k'}}W_{
e-e}({k},{k'})
\end{equation}
and
 to the renormalization of the Rabi frequency $\Omega_{
k}(t)=i\frac{e_0}{m_0}M^{cv}_z(k)A_z(t)$ due to the attractive
electron-hole interaction $W_{e-h}({k},
{k'})$
\begin{equation}
\label{exc}
 \tilde{\Omega}_{
k}(t)=\Omega_{
k}(t)+\frac{1}{\hbar
}\sum_{k'}W_{e-h}({k}, {k'}) 
p_{k'}(t)\,.
\end{equation}
Here, 
$M^{cv}_z(k)$ is the optical matrix element describing the strength of the carrier-light coupling and $A_z(t)$ is the vector potential denoting the optical excitation of the system with light polarized along the nanotube axis (here, z-axis).   We obtain an analytic expression for $M^{cv}_z(k)$ within the nearest-neighbor tight-binding approximation \cite{malic06}. The screened Coulomb matrix elements 
$W^{v{k}',v{k}}_{v{k},v{k}'}$ determining $W_{e-e}({k},{k'})$ and $W^{v{k}',v{k}}_{v{k},v{k}'}$ in Eqs. (\ref{ren}), (\ref{exc})
\cite{malic10} are obtained by explicitly calculating the appearing tight-binding coefficients and by applying the Ohno parametrization of the Coulomb potential, which is known to be a good approximation for CNTs \cite{jiang07}. 

\begin{figure}[h!]
	\centering
	\includegraphics[width=0.50\textwidth]{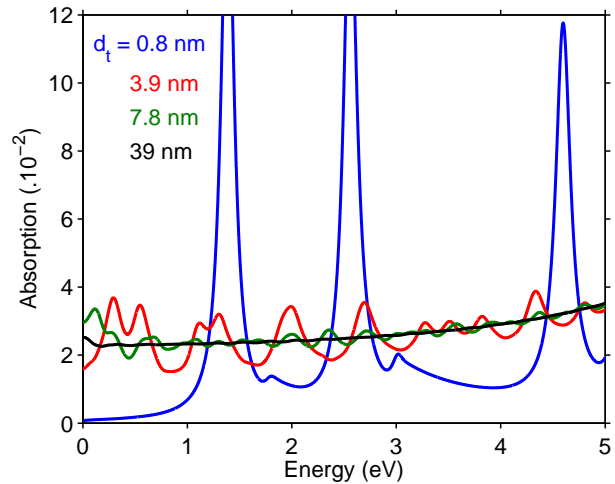}
	\caption{Calculated absorption of nanotubes for selected diameters. For larger diameters, it clearly appears that the absorption becomes diameter independent and reaches the one of graphene. The calculations are normalized so as to set the value of this plateau to $\pi \alpha$. \label{fig:calcgraphene}}
\end{figure}

By numerically evaluating the Bloch equation for the microscopic polarization $p_k(t)$, we have access to the absorption of carbon nanotubes with arbitrary chiral angle and with a wide range of diameters. Here, we calculated the absorption coefficient of the investigated (6,5) and (10,2) nanotubes representating near armchair and near zigzag species (Figure~\ref{fig:PLEspectrum}). The absolute value of the absorption was obtained by scaling the graphene absorption calculated in the same way to $\pi\alpha$ \cite{Nair2008}, \footnote{The effect of external dielectric screening would lead to a slight decrease of this value of the order of 10 to 15\%. However, due to the poor knowledge on the nano-environment of the tube in the micelle structure we neglected this contribution.}. We further checked that the calculated absorption of very large nanotubes tends to the graphene limit (Figure~\ref{fig:calcgraphene}). In order to match the experimentally measured $S_{22}$ resonance, we applied a global energy shift 
to the theoretical spectra to account for the environment-induced screening of the Coulomb potential \cite{Berger2009} and a broadening of $\simeq$150~meV to account for the many-particle-induced broadening that has not been considered on the applied Hartree-Fock level. Note that the broadening does not alter the non-resonant absorption value since the variations of the latter are very small at this scale (see Inset of Fig.~\ref{fig:PLEspectrum}). The agreement between theory and experiment is remarkable regarding this non-resonant absorption in the sense that there are no free parameters in the theory. In particular, both the amplitude and the profile of the non-resonant spectrum are very well reproduced. In contrast, the excitonic resonances significantly differ, which can probably be traced back to the presence of phonon-induced and other side-bands that have not been taken into account in the theory and that withdraw parts of the oscillator strength from the main excitonic transition.

\begin{figure}[t!]
	\centering
	\includegraphics[width=0.50\textwidth]{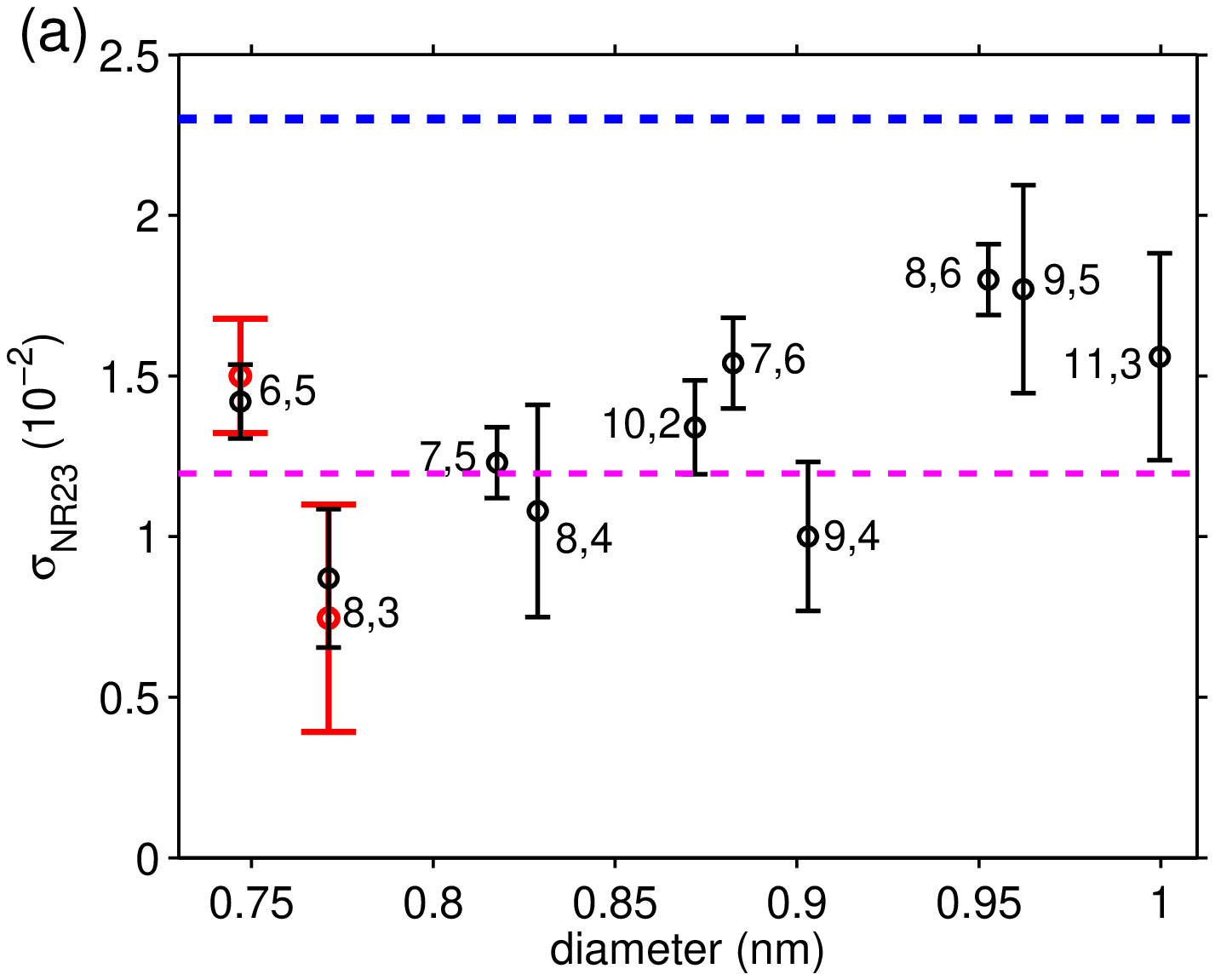}
	\includegraphics[width=0.5\textwidth]{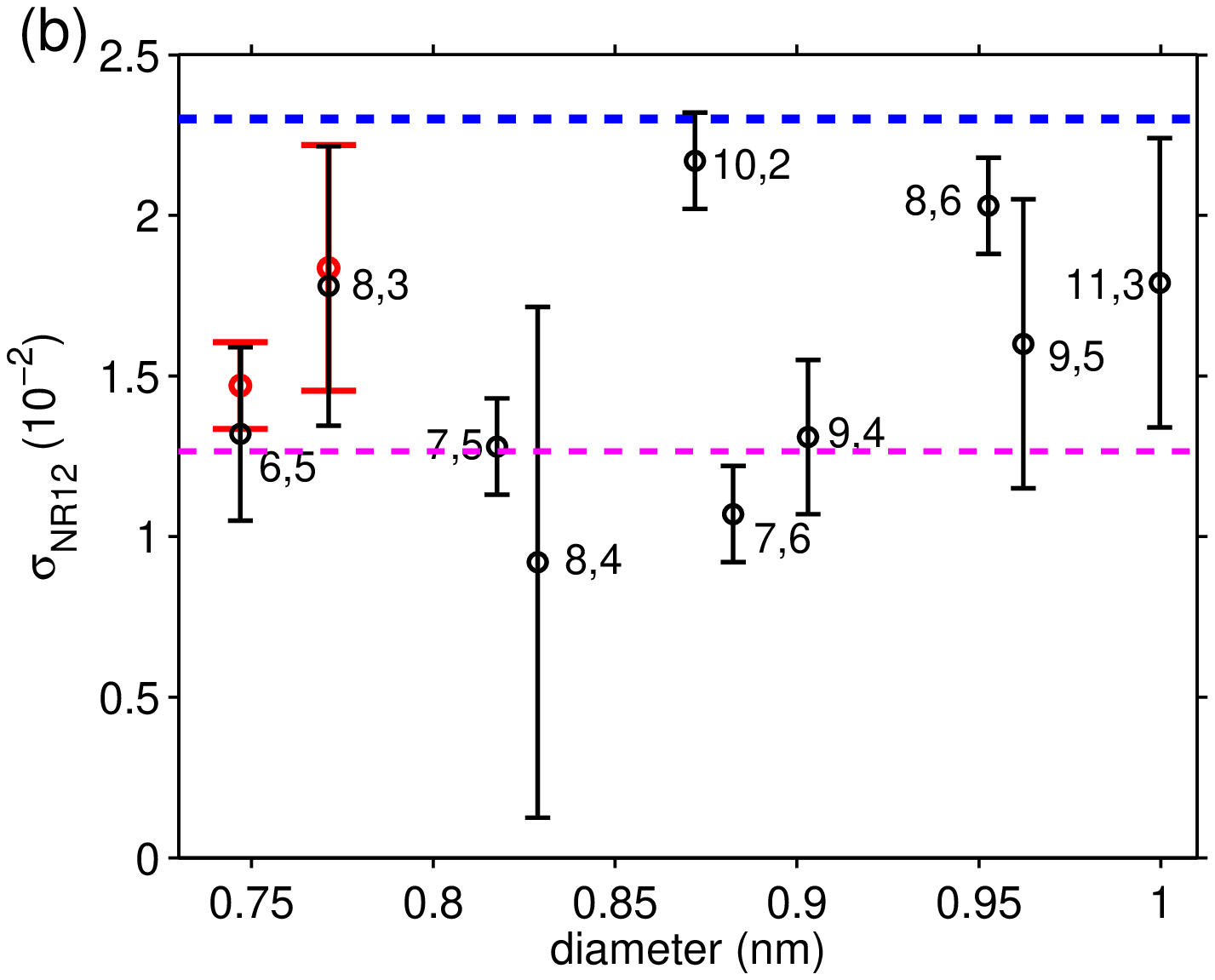}
	\caption{Diameter dependence of the non-resonant absorption contribution extracted from deconvoluted PLE spectra for 10 species either between the $S_{22}$ and $S_{33}$ ($\sigma_{\text{NR}23}$) (a) or the $S_{11}$ and $S_{22}$ transitions ($\sigma_{\text{NR}12}$) (b). The black symbols stand for the HiPCO material and the red ones for the CoMoCat material. The blue dashed line represents the graphene absorption, while the magenta dashed line shows non-resonant absorption of nanotubes estimated within the zone-folding approximation. \label{fig:HRSvsD}}	
\end{figure}

The experimental non-resonant absorption contribution was extracted for 10 species (Figure~\ref{fig:HRSvsD}). This absorption shows no clear dependence on the diameter of the species, with overall variations smaller than the mean error bar. We also examinated a possible dependence on the chiral angle but could not see any effect (Appendix, Figure~\ref{fig:angle}). This is in strong contrast to the case of the resonant $S_{22}$ transition that shows a remarkable chiral angle dependence \cite{Vialla2013, verdenhalven13}. Notably, the non-resonant absorption cross-section is almost identical between the $S_{11}$ and $S_{22}$ resonances or between the $S_{22}$ and $S_{33}$ resonances and corresponds to approximately one half of the value for graphene.

Interestingly, most of these observations -including the non-resonant absorption of approximately one half of that of graphene- can be qualitatively understood in the framework of a simple band-to-band absorption theory. Actually, the non-resonant contribution is observed for energies well above the excitonic resonance, where Coulomb correlation corrections are expected to be weak. In fact, this independent electron description allowed to compute quite accurately the graphene absorption in the NIR and visible range \cite{Nair2008, Mak2008}. Let us first note that from a simple oscillator strength conservation argument, the background absorption of nanotubes is expected to be smaller than that of graphene because most of the oscillator strength is transferred into the resonances of the nanotubes. More quantitatively, the absorption probability is given by the Fermi golden rule and involves the dipolar matrix elements and the joint density of states (JDOS) between the valence and conduction sub-bands. In a first approximation, the latter can be obtained from a zone-folding approach using the conical approximation for the graphene band-structure in the vicinity of the $K$ points. The JDOS of a nanotube is computed per graphene unit cell (or equivalently per two carbon atoms) \cite{Mintmire1998}:
\begin{equation}
  j_{NT}(\omega)= \frac{\sqrt{3} a^2}{2 \pi^2 \hbar v_F d_t} \sum_i g(\hbar \omega, S_{ii})
  \label{eq:DOS}
\end{equation}
where \begin{equation*}
       g(\hbar \omega, E_i) = \begin{cases} |\hbar \omega|/\sqrt{(\hbar \omega)^2-E_i^2}  & \mbox{if } |\hbar \omega| > E_i \\
                     0  & \mbox{if } |\hbar \omega| < E_i \end{cases}
      \end{equation*}

where $a$ is the length of the graphene lattice base vector, $v_F=10^6$~m.s$^{-1}$ is the graphene Fermi velocity and $d_t$ is the nanotube diameter.
Using the same notations, the joint density of states per unit cell for graphene reads:
\begin{equation}
 j_G(\omega)=\frac{\sqrt{3}a^2}{8\pi(\hbar v_F)^2}\hbar \omega
\end{equation}

 The key point here is that $j_{NT}(\omega)$ tends to a plateau when the photon energy is much larger than the excitonic energy and increases stepwise for each additional $S_{ii}$ transition. This is in contrast to the case of graphene where the joint density of states grows linearly with the photon energy, leading to the well known frequency independent absorption probability. The link between the 2D graphene case and its 1D CNT counterpart  lies in the number of plateaus that fit within the $\hbar \omega$ energy window. 
 Therefore, there is a profound relationship between the absorption probability of a nanotube away from the resonances and that of a graphene layer. 
 In a first approximation, we neglected the possible variations of the matrix element with the direction of electron wave-vector. Due to the overall frequency dependence of the optical matrix element and of the prefactors \cite{grueneis03,malic06}, the JDOS plateaus turn into slowly decreasing tails in the absorption spectrum 
in agreement with the experimental observations and the microscopic calculations (see Figure~\ref{fig:PLEspectrum}).
  

More quantitatively, this simple model allows us to compare the JDOS per carbon atom (and therefore the absorption) in the cases of graphene and of nanotubes at a photon energy at the foot of the $S_{ii}$ resonances of the nanotubes (where the non-resonant contribution is most readily observable (Figure~\ref{fig:PLEspectrum})). One finds $j_{NT}/j_G = \sqrt{3}/\pi \simeq 0.55$ at the foot of the $S_{22}$ transition and $j_{NT}/j_G = \frac{3}{4\pi} (\frac{4}{\sqrt{15}} +\frac{2}{\sqrt{3}}) \simeq 0.52$ at the foot of the $S_{33}$ transition, leading to an absorption of $\simeq 0.012$ in the nanotube case, in good agreement with the mean experimental value $0.013 \pm 0.003$ (Figure~\ref{fig:HRSvsD}).

\begin{figure}[t!]
	\centering
	\includegraphics[width=0.5\textwidth]{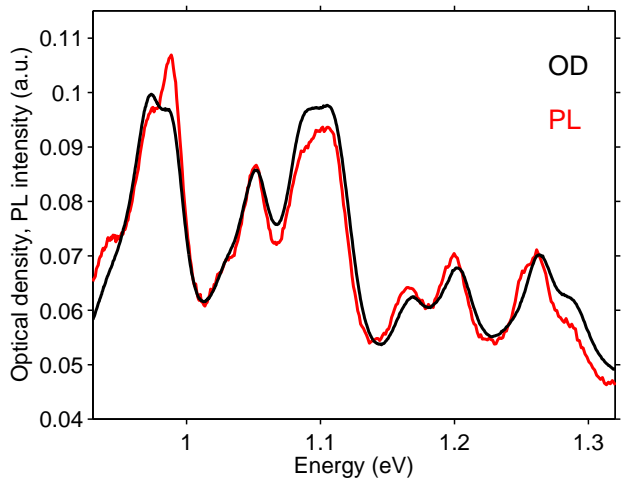}
	\caption{Absorption spectrum (black line) of a sodium cholate suspension of HiPCO nanotubes (offset corrected and normalized). PL spectrum (red line) of the same sample for an off-excitation wavelength at 3~eV. \label{fig:abs-vs-PL}}	
\end{figure}

Incidentally, this universal off-resonant absorption (that depends neither on the frequency, nor on the transition order, nor on the chiral species) explains why non-resonantly excited photo-luminescence spectra are empirically known to give a good image of the chiral species content of a sample. This effect simply results from the fact that such a non-resonant excitation provides a uniform excitation of the sample at a given wavelength regardless of the chiral species. This effect is depicted in Figure~\ref{fig:abs-vs-PL} where the linear absorption spectrum of a HiPCO sample is compared to its PL spectrum excited off-resonance (for a laser wavelength falling inbetween the $S_{22}$ and $S_{33}$ resonances of most chiral species). It is clearly observable that the magnitude of the lines observed in the PL spectrum is in good agreement with that of the absorption spectrum.

In conclusion, we used PLE measurements with global fitting analysis to delineate the contribution of each $(n,m)$ species to the non-resonant absorption background of a micelle suspension of SWNTs. We show that this non-resonant absorption reaches an universal value for all  chiral species, equal to approximately one half of the value of the unrolled graphene sheet both between the $S_{11}$ and $S_{22}$ and between the $S_{22}$ and $S_{33}$ resonances. In addition, this non-resonant absorption hardly depends on the photon energy over 0.5~eV wide windows inbetween excitonic resonances. These properties of the non-resonant absorption of SWNT were quantitatively reproduced by microscopic calculations based on the density matrix formalism. This study has practical implications for non conventional excitation schemes of ensembles of SWNTs. Furthermore, it exemplifies the profound link between the photophysical properties of the differents nanocarbons.

\clearpage

\newpage

\section{Appendix}

\subsection{Time-resolved measurements}

\begin{figure}[h!]
	\centering
	\includegraphics[width=0.50\textwidth]{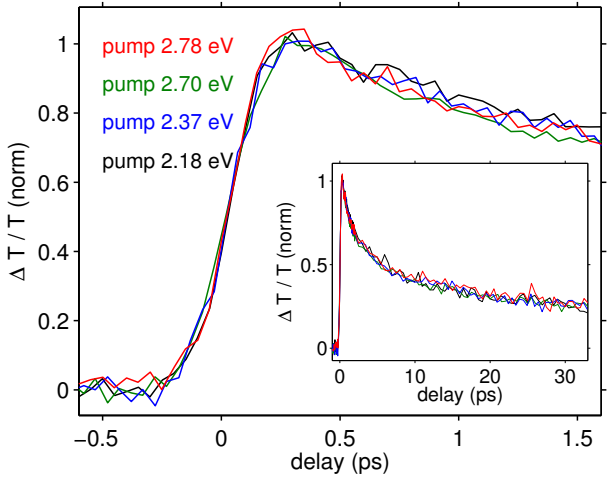}
	\caption{Transient bleaching of the $S_{11}$ transition of (6,5) nanotubes in a micellar suspension subsequent to femto-second excitation at either their $S_{22}$ transition (black line) or at several non-resonant energies as indicated in the figure. In all cases, the product of the pump power density with the absorbance was kept constant resulting in similar excitation densities. Inset : same data plotted on an extended time window. \label{fig:femto}}
\end{figure}

Femtosecond pump-probe measurements were carried out to probe the population build-up dynamics of the $S_{11}$ level subsequent to either $S_{22}$ or non-resonant excitation (Figure~\ref{fig:femto}). The strictly identical rise-times of the transients (within our 250~fs time-resolution) are consistent with ultrafast internal conversion, making possible a direct scaling of PLE spectra into absorption.

\subsection{Temperature dependence}

We measured the PLE spectrum of an ensemble of nanotubes deposited on a quartz plate as a function of temperature. After normalization, the PLE profiles are identical showing that the internal relaxation towards the $S_{11}$ level is much faster than any other competing process. In fact, if this was not the case, the temperature-induced changes expected for these processes would warp the PLE profile.

\subsection{Band to band calculations}

\begin{figure}[t!]
	\centering
	\includegraphics[width=0.50\textwidth]{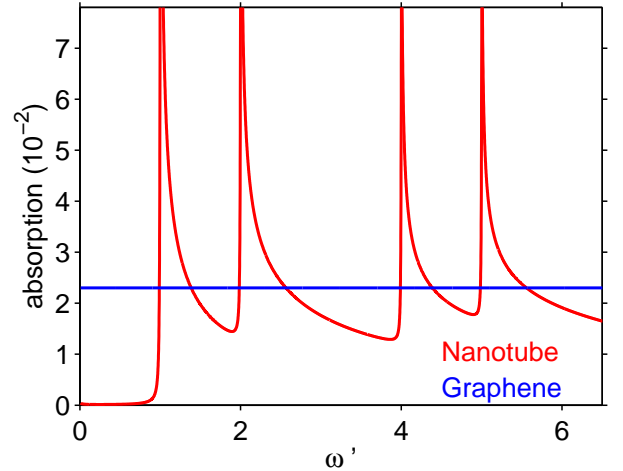}
	\caption{Absorption calculated from the joint density of states in the conical zone-folding approximation (red line) as a function of the reduced energy $ \omega'=3 \omega d_t /(4 v_F)$, where $v_F \simeq 1\times 10^6$~m.s$^{-1}$ stands for the graphene Fermi velocity. The amplitude is rescaled globally so that the value computed for a graphene layer matches the universal value of 0.023 (blue solid line). An effective broadening of 0.01~eV was used. \label{fig:DOS}}
\end{figure}

Figure~\ref{fig:DOS} shows the absorption spectrum calculated in the band-to-band approximation both for graphene and nanotubes as a function of reduced frequency $ \omega'=3 \omega d_t /(4 v_F)$, where $v_F \simeq 1\times 10^6$~m.s$^{-1}$ stands for the graphene Fermi velocity. The JDOS plateaus of the nanotube give rise to slowly decreasing absorption tails between the resonances, reaching approximately one half of the absorption of graphene at the foot of each resonance.

\subsection{Global analysis method}

\begin{figure}[h!]
	\centering
	\includegraphics[width=0.50\textwidth]{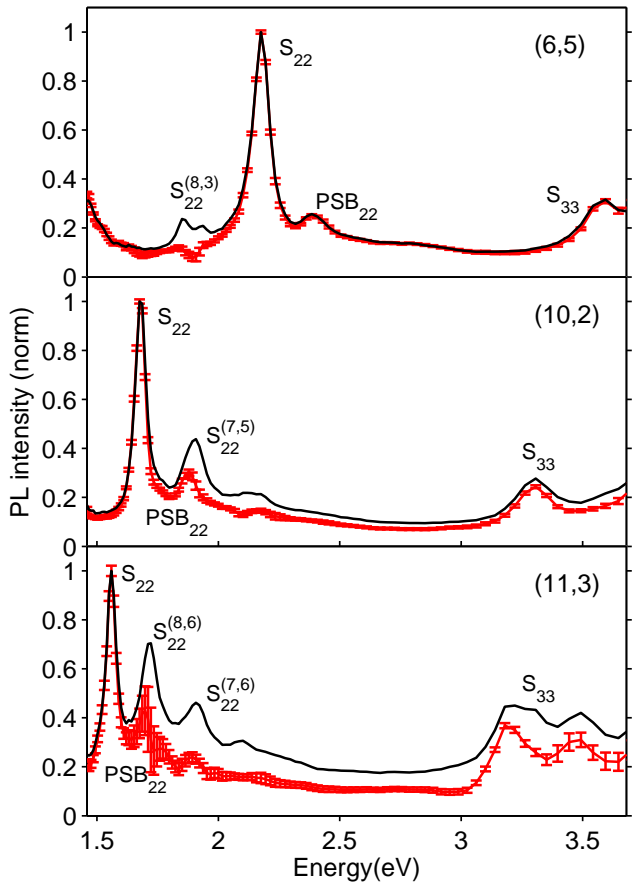}
	\caption{Comparison of regular PLE spectra (black solid line) with deconvoluted spectra (red line) from the global analysis method for selected chiral species with associated error bars. \label{fig:globalanalysis}}
\end{figure}

 The efficiency of the global analysis method is exemplified in Figure~\ref{fig:globalanalysis} where the contribution of minority species showing up as side peaks in regular PLE spectra are efficiently suppressed in the deconvoluted PLE spectrum.
 
\subsection{Chiral angle dependence}

\begin{figure}[h!]
	\centering
	\includegraphics[width=0.50\textwidth]{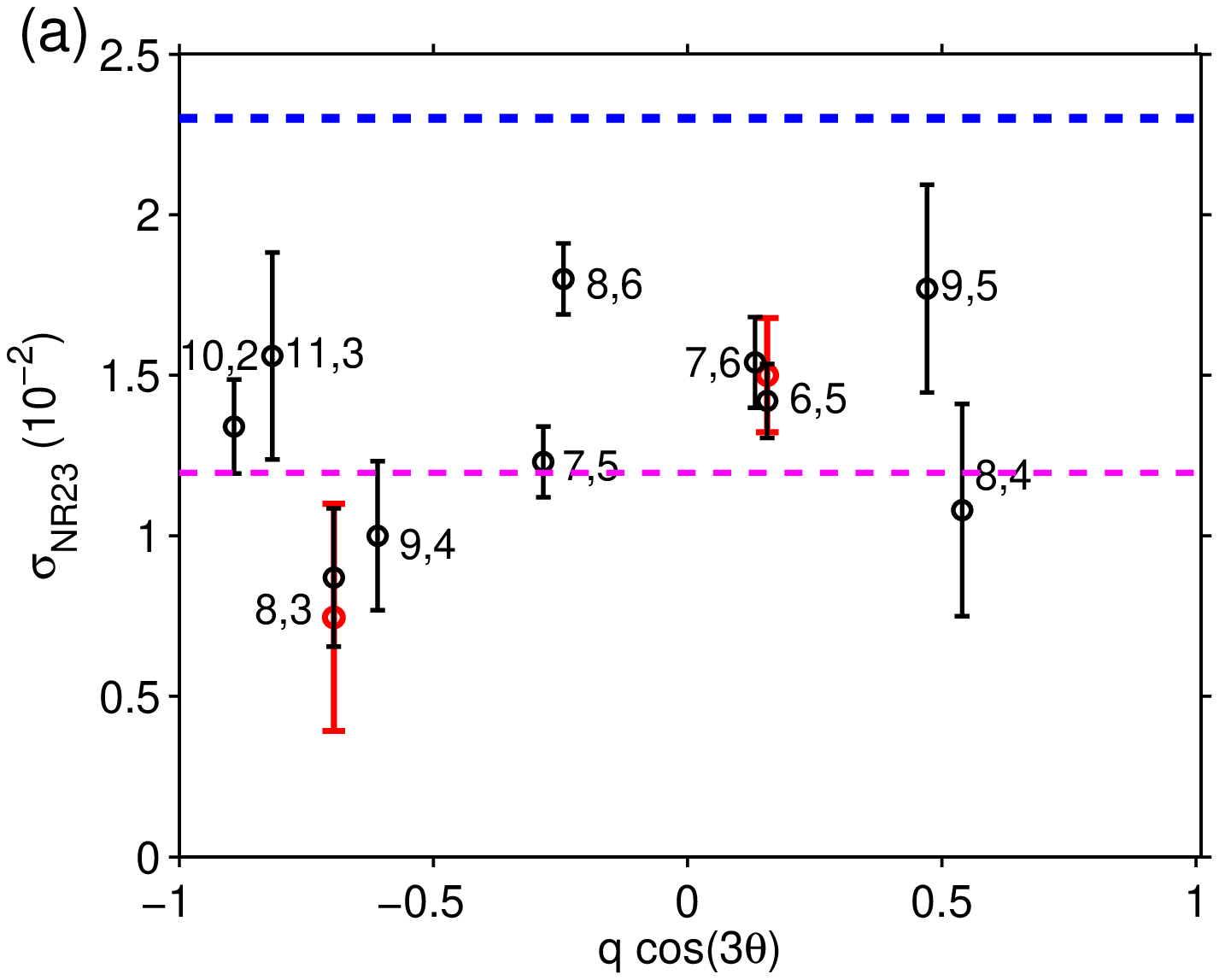}
	\includegraphics[width=0.50\textwidth]{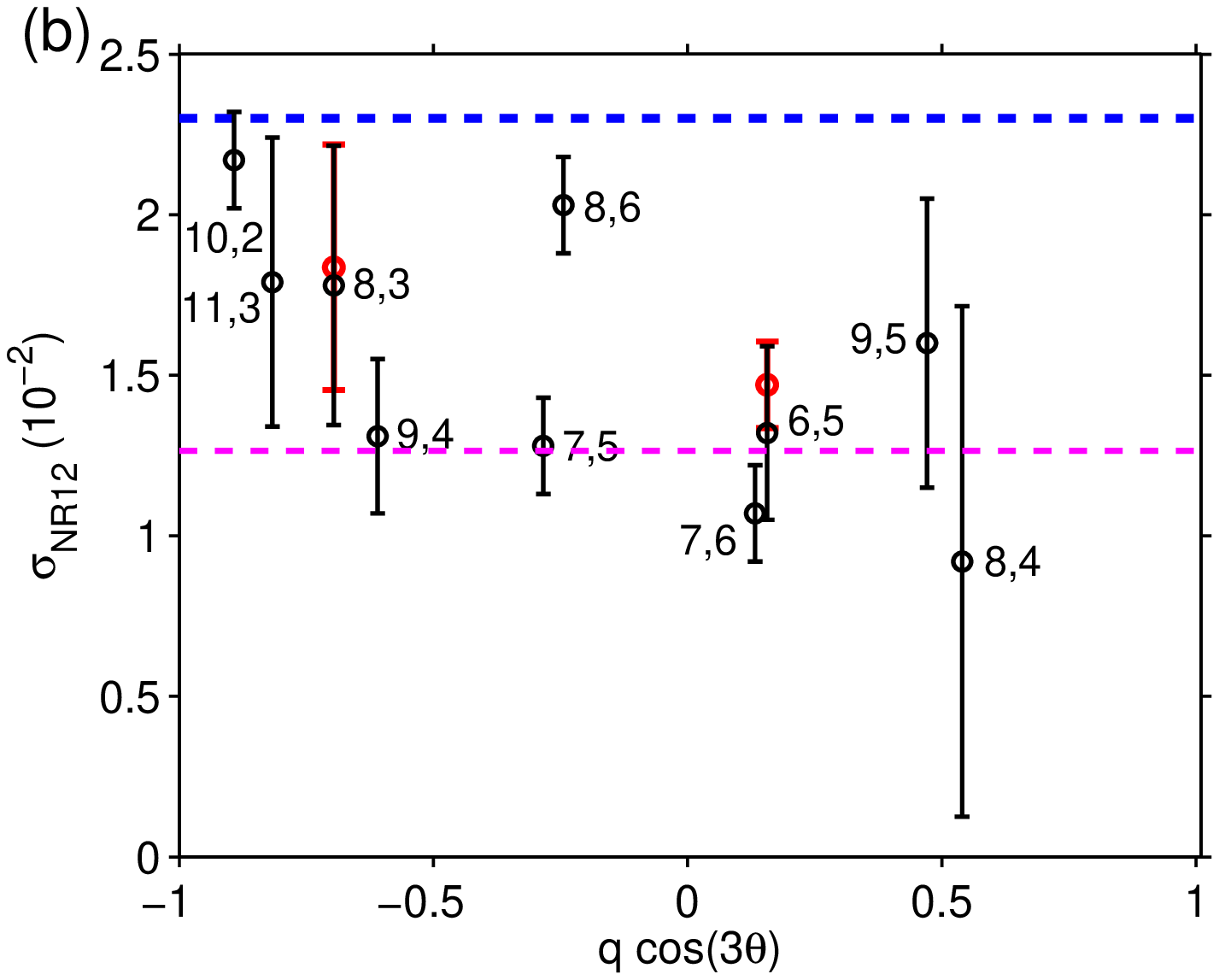}
	\caption{Non-resonant absorption of several chiral species measured at the foot of their $S_{33}$ (a) and $S_{22}$ (b) transition as a function of the trigonal parameter $q \cos 3 \theta$, where $q = n-m$~mod~3 . \label{fig:angle}}	
\end{figure}

The non-resonant absorption cross-section extracted for 10 chiral species can be plotted as a function of their chiral angle (Figure~\ref{fig:angle}). Following previous studies \cite{Vialla2013}, we rather plot them as a function of the trigonal parameter $q \cos 3 \theta$ (where $q = n-m$~mod~3) which is more suited to describe chiral variations of the electronic properties in SWNTs. Within our error bars, we do not observe any systematic variation of the non-resonant absorption with the chiral angle. However, a nascent trend can be guessed that would correspond to a decrease of $\sigma_{\text{NR}12}$ with increasing $q \cos 3 \theta$ and a symmetrical increase of $\sigma_{\text{NR}23}$. We believe that these are not intrinsic features but that this rather arises from the decrease with increasing $q \cos 3 \theta$ of the energy splitting between the $S_{22}$ and $S_{33}$ resonances as a consequence of the trigonal warping (and inversely for $S_{11}$ and $S_{22}$). Therefore, the background is slightly 
overestimated 
for $\sigma_{\text{NR}23}$ due to overlapping with the tails of the resonances for positive $q \cos 3 \theta$ values (and inversely for $\sigma_{\text{NR}12}$).

%
%

\subsection*{Acknowledgement}
This work was supported by the
GDR-I GNT and the ANR grant "TRANCHANT". CV is a member of ``Institut Universitaire de France''.  E.M. acknowledges financial support from the Einstein Foundation Berlin.\\

\bibliography{biblio-sigma-NR}

\end{document}